\documentstyle[12pt]{article}
\renewcommand{\baselinestretch}{2.0}

\begin{document}
{\Large
\centerline {
On the sensitivity of wave channeling of X-ray beam
}
\centerline {
to the shape of interface channels.
}
}
\vskip 15pt
%\author
{
\centerline {T.A.Bobrova, L.I.Ognev\footnote[1]{
Nuclear Fusion Institute,
Russian Research Center "Kurchatov
Institute", Moscow, 123182, Russia \\
E-mail: ognev@nfi.kiae.ru
}
}
\par
{\abstract
%\par
The using of microdiffraction of X-ray radiation for analysis of the
structure
of material specimens with submicron resolution becomes very 
promising
investigation method \cite{Riekel2000}.
One of the methods for obtaining of submicron beams of hard X-ray 
radiation
is formation in a narrow channel of dielectrical resonator 
\cite{Riekel2000, Jark1999p9}.
In the present work
the effect of transmission of X-ray through narrow submicron rough
channels
was investigated by numerical simulation with account for 
diffraction
and decay of coherency. It was found that transmission
can be strongly decreased for channels with periodic deformations.
The effects of roughnes were explained
with the statistical theory of X-ray scattering in rough transitional 
layer.
The wave mode attenuation coefficients $\beta$ scale as
$\beta \sim {1/{d^3}}$ ($d$ is channel width)
and proportional to roughness amplitude $\sigma$.
Possible explanation of observed anomalous energy
dependence of transmission through thin Cr/C/Cr channel
was given. The sensitivity of transmission of dielectrical channel 
to the presence
of roughness and deformation with large space period was 
investigated.
}
%\vskip 25pt

\newpage

The using of microdiffraction of X-ray radiation for analysis of the
structure
of material specimens with submicron resolution becomes very 
promising
investigation method \cite{Riekel2000}.
One of the methods for obtaining of submicron beams of hard X-ray 
radiation
is formation in a narrow channel of dielectrical resonator 
\cite{Riekel2000, Jark1999p9}.
Monitoring of X-ray beam by capture into a narrow dielectric 
channel
is used in waveguide X-ray laser physics 
\cite{Kukhlevsky1999pure, Kukhlevsky2000}, 
production of thin X-ray
probe beams \cite{Jark1999p9} and other applications 
\cite{Kantsyrev1995}
due to the effect of
total external reflection.

%Minimum angle of divirgence of the beam in this case is limited 
%by
%Fresnel angle $\vartheta_F$ and rougness scattering can only 
%deteriorate
%this parameter.

In this work we consider the role of diffraction that can
be important for narrow beams especielly when roughness is high.
Scattering from surfaces with high roughness needs special 
approach
because small perturbation methods fail \cite{Fanchenko1999}.

\centerline{\bf Theoretical model}
X-ray scattering on rough surfaces is usually investigated within
the well known Andronov-Leontovich approach 
\cite{Vinogradov1985}
but for very small angles of incidence
the model of "parabolic equation" for slowly varying scalar
amplitudes of electrical field vector $A(x,z)$
should be used.
Within the  model scattering and absorption do not disappear
at small grazing angle limit that results from
Andronov-Leontovich approach \cite{Vinogradov1985}.
In this case large angle scattering is neglected so
$$
\partial^2 A(x,z)/ \partial z^2 \ll k \cdot \partial A(x,z)/ 
\partial z
$$
and because  the beam is narrow
$$
\partial^2 A(x,z)/ \partial z^2 \ll
\partial^2 A(x,z)/ \partial x^2 ,
$$
where $z$ and $x$ are coordinates along and across the channel.
The consideration will be restricted here to 2-dimensional channels
(gaps) although the same approach can be applied to capillaries.
The assumption results in "parabolic equation" of quazioptics
\cite{BobrovaJEPT1999}:
%%%%%%%%%%%%%%%%%%%%%%%%%%%%%%%%%
%%%%%%%%%%%%%%%%%%%%%%%%%%%
%$\partial ^{2}A/\partial z^{2}  \ll k(\partial A/\partial z)$
%and
%$\partial ^{2}A/\partial x^{2}$:
\par
$$
2ik {\partial  A\over \partial  z} =
\Delta _{\perp }A + k^{2} {{\varepsilon- \varepsilon_0} \over
\varepsilon _{0}} A
\eqno(1)
$$
\noindent
$$
A(x,z=0)=A_{0}(x),
$$
\noindent where
%$z$ is coordinate along beam direction,
$k = \sqrt{\varepsilon} _{0}{\omega \over c}$.
(In this case $\varepsilon_0$ is dielectrical permittance of air,
$\varepsilon_1$ - dielectrical permittance of glass.)
The evolution of the channeled X-ray beam
%can
%be
was calculated by direct integration of the "parabolic" equation
\cite{BobrovaPhysStatSol1997}.
The dielectric permitance on the rough boundary with the random 
shape
$x=\xi (z)$ was presented as
\par
\noindent $\varepsilon(x,z)=\varepsilon_1
+(\varepsilon_0 - \varepsilon_1 )H(x-\xi(z))
$
where
%$\varepsilon_1$ and $\varepsilon_0$ is the dielectric 
%permittance
%of the substance and of the air respectively,
$H(x)$ is a step function.
The distribution of roughness heights
is assumed to be normal.
It is known from results of \cite{Vinogradov1985} that at grazing 
incidence
the effect of scatering is very small. So special surfaces are
needed to observe scattering effects in the gap interface at
reasonable distance. In the calculations we used roughness 
amplitude
up to $400 \AA$.
The reflection of X-ray beam on very rough surfaces (up to 1500 \AA)
of silicon was observed in\cite{Tsuji1995}.
The results of direct simulation of scattering with the model rough 
surface
by integration of equation (1) calculated for X-ray
energy $E=10 keV$, width of the channel $d=0.5 \mu m$,
$\sigma=400 \AA$ and correlation length of roughness
$z_{corr}=5\mu m$ averaged over 40 realizations are shown on 
Fig.1
as normalized to initial value total intensity of the beam $r_{tot}$,
incoherent part $r_{inc}$, where
$r_i=\int_{-\infty}^{\infty}I_i(x)dx\slash 
\int_{-d/2}^{d/2}I_0(x)dx$.
Initial angles of incidence of plane wave were
$\vartheta=0$; $3\cdot 10^{-4}$
and $6\cdot 10^{-4} rad$ (Fresnel angle 
$\vartheta_F = 3\cdot 10^{-3}rad$).
%%%%%%%%%%%%%%%%%%%%%%%%%%%%%%%%
The atomic scattering factors used in the calculations were taken from
\cite{Henke1993}.
%\vskip 20pt
\begin{center}
\begin{minipage}{11cm}
% GNUPLOT: LaTeX picture with emtex specials

%\input {fig1zhet.tex}
% GNUPLOT: LaTeX picture with emtex specials
\setlength{\unitlength}{0.240900pt}
\ifx\plotpoint\undefined\newsavebox{\plotpoint}\fi
\sbox{\plotpoint}{\rule[-0.200pt]{0.400pt}{0.400pt}}%
\special{em:linewidth 0.4pt}%
\begin{picture}(1500,900)(0,0)
%\begin{picture}(250,150)(0,0)
\font\gnuplot=cmr10 at 10pt
\gnuplot
\put(220,113){\special{em:moveto}}
\put(1436,113){\special{em:lineto}}
\put(220,113){\special{em:moveto}}
\put(220,832){\special{em:lineto}}
\put(220,113){\special{em:moveto}}
\put(240,113){\special{em:lineto}}
\put(1436,113){\special{em:moveto}}
\put(1416,113){\special{em:lineto}}
\put(198,113){\makebox(0,0)[r]{0}}
\put(220,233){\special{em:moveto}}
\put(240,233){\special{em:lineto}}
\put(1436,233){\special{em:moveto}}
\put(1416,233){\special{em:lineto}}
\put(198,233){\makebox(0,0)[r]{0.2}}
\put(220,353){\special{em:moveto}}
\put(240,353){\special{em:lineto}}
\put(1436,353){\special{em:moveto}}
\put(1416,353){\special{em:lineto}}
\put(198,353){\makebox(0,0)[r]{0.4}}
\put(220,473){\special{em:moveto}}
\put(240,473){\special{em:lineto}}
\put(1436,473){\special{em:moveto}}
\put(1416,473){\special{em:lineto}}
\put(198,473){\makebox(0,0)[r]{0.6}}
\put(220,592){\special{em:moveto}}
\put(240,592){\special{em:lineto}}
\put(1436,592){\special{em:moveto}}
\put(1416,592){\special{em:lineto}}
\put(198,592){\makebox(0,0)[r]{0.8}}
\put(220,712){\special{em:moveto}}
\put(240,712){\special{em:lineto}}
\put(1436,712){\special{em:moveto}}
\put(1416,712){\special{em:lineto}}
\put(198,712){\makebox(0,0)[r]{1}}
\put(220,832){\special{em:moveto}}
\put(240,832){\special{em:lineto}}
\put(1436,832){\special{em:moveto}}
\put(1416,832){\special{em:lineto}}
\put(198,832){\makebox(0,0)[r]{1.2}}
\put(220,113){\special{em:moveto}}
\put(220,133){\special{em:lineto}}
\put(220,832){\special{em:moveto}}
\put(220,812){\special{em:lineto}}
\put(220,68){\makebox(0,0){0}}
\put(524,113){\special{em:moveto}}
\put(524,133){\special{em:lineto}}
\put(524,832){\special{em:moveto}}
\put(524,812){\special{em:lineto}}
\put(524,68){\makebox(0,0){5000}}
\put(828,113){\special{em:moveto}}
\put(828,133){\special{em:lineto}}
\put(828,832){\special{em:moveto}}
\put(828,812){\special{em:lineto}}
\put(828,68){\makebox(0,0){10000}}
\put(1132,113){\special{em:moveto}}
\put(1132,133){\special{em:lineto}}
\put(1132,832){\special{em:moveto}}
\put(1132,812){\special{em:lineto}}
\put(1132,68){\makebox(0,0){15000}}
\put(1436,113){\special{em:moveto}}
\put(1436,133){\special{em:lineto}}
\put(1436,832){\special{em:moveto}}
\put(1436,812){\special{em:lineto}}
\put(1436,68){\makebox(0,0){20000}}
\put(220,113){\special{em:moveto}}
\put(1436,113){\special{em:lineto}}
\put(1436,832){\special{em:lineto}}
\put(220,832){\special{em:lineto}}
\put(220,113){\special{em:lineto}}
\put(45,472){\makebox(0,0){$r_i$}}
\put(828,23){\makebox(0,0){$\mu m$}}
\put(828,877){\makebox(0,0)}
\put(646,413){\makebox(0,0){$1$}}
\put(646,293){\makebox(0,0){$1'$}}
\put(445,365){\makebox(0,0)[l]{$2$}}
\put(402,443){\makebox(0,0)[l]{$2'$}}
\put(342,413){\makebox(0,0)[l]{$3$}}
\put(342,652){\makebox(0,0)[l]{$3'$}}
\sbox{\plotpoint}{\rule[-0.600pt]{1.200pt}{1.200pt}}%
\special{em:linewidth 1.2pt}%
\put(248,832){\special{em:moveto}}
\put(256,767){\special{em:lineto}}
\put(258,760){\special{em:lineto}}
\put(296,630){\special{em:lineto}}
\put(334,570){\special{em:lineto}}
\put(372,528){\special{em:lineto}}
\put(524,426){\special{em:lineto}}
\put(676,354){\special{em:lineto}}
\put(828,303){\special{em:lineto}}
\put(980,264){\special{em:lineto}}
\put(1132,231){\special{em:lineto}}
\put(1284,205){\special{em:lineto}}
\put(1436,183){\special{em:lineto}}
\sbox{\plotpoint}{\rule[-0.500pt]{1.000pt}{1.000pt}}%
\special{em:linewidth 1.0pt}%
\put(229,283){\usebox{\plotpoint}}
\multiput(229,283)(7.288,19.434){2}{\usebox{\plotpoint}}
\multiput(238,307)(15.513,13.789){0}{\usebox{\plotpoint}}
\put(250.81,314.58){\usebox{\plotpoint}}
\multiput(256,314)(18.564,-9.282){0}{\usebox{\plotpoint}}
\multiput(258,313)(17.549,-11.083){2}{\usebox{\plotpoint}}
\multiput(296,289)(20.072,-5.282){2}{\usebox{\plotpoint}}
\multiput(334,279)(19.306,-7.621){2}{\usebox{\plotpoint}}
\multiput(372,264)(20.611,-2.441){7}{\usebox{\plotpoint}}
\multiput(524,246)(20.739,0.819){8}{\usebox{\plotpoint}}
\multiput(676,252)(20.739,0.819){7}{\usebox{\plotpoint}}
\multiput(828,258)(20.754,0.273){7}{\usebox{\plotpoint}}
\multiput(980,260)(20.668,1.904){8}{\usebox{\plotpoint}}
\multiput(1132,274)(20.755,0.137){7}{\usebox{\plotpoint}}
\multiput(1284,275)(20.719,1.227){7}{\usebox{\plotpoint}}
\put(1436,284){\usebox{\plotpoint}}
\sbox{\plotpoint}{\rule[-0.200pt]{0.400pt}{0.400pt}}%
\special{em:linewidth 0.4pt}%
\put(249,832){\special{em:moveto}}
\put(272,681){\special{em:lineto}}
\put(298,592){\special{em:lineto}}
\put(323,530){\special{em:lineto}}
\put(350,478){\special{em:lineto}}
\put(375,433){\special{em:lineto}}
\put(401,393){\special{em:lineto}}
\put(427,360){\special{em:lineto}}
\sbox{\plotpoint}{\rule[-0.500pt]{1.000pt}{1.000pt}}%
\special{em:linewidth 1.0pt}%
\put(246,338){\usebox{\plotpoint}}
\multiput(246,338)(6.006,19.867){5}{\usebox{\plotpoint}}
\put(285.32,428.10){\usebox{\plotpoint}}
\put(305.13,434.28){\usebox{\plotpoint}}
\multiput(323,440)(13.338,15.903){2}{\usebox{\plotpoint}}
\multiput(349,471)(20.740,-0.798){2}{\usebox{\plotpoint}}
\put(392.72,475.45){\usebox{\plotpoint}}
\put(413.09,478.00){\usebox{\plotpoint}}
\put(427,478){\usebox{\plotpoint}}
\sbox{\plotpoint}{\rule[-0.200pt]{0.400pt}{0.400pt}}%
\special{em:linewidth 0.4pt}%
\put(247,832){\special{em:moveto}}
\put(259,723){\special{em:lineto}}
\put(272,637){\special{em:lineto}}
\put(284,571){\special{em:lineto}}
\put(298,513){\special{em:lineto}}
\put(311,465){\special{em:lineto}}
\put(323,426){\special{em:lineto}}
\sbox{\plotpoint}{\rule[-0.500pt]{1.000pt}{1.000pt}}%
\special{em:linewidth 1.0pt}%
\put(233,349){\usebox{\plotpoint}}
\multiput(233,349)(2.050,20.654){7}{\usebox{\plotpoint}}
\multiput(246,480)(2.873,20.556){4}{\usebox{\plotpoint}}
\put(261.69,573.62){\usebox{\plotpoint}}
\multiput(272,576)(7.413,19.387){2}{\usebox{\plotpoint}}
\put(292.78,623.16){\usebox{\plotpoint}}
\put(304.83,639.96){\usebox{\plotpoint}}
\put(318.71,655.38){\usebox{\plotpoint}}
\end{picture}

%\vskip 15pt
%\par
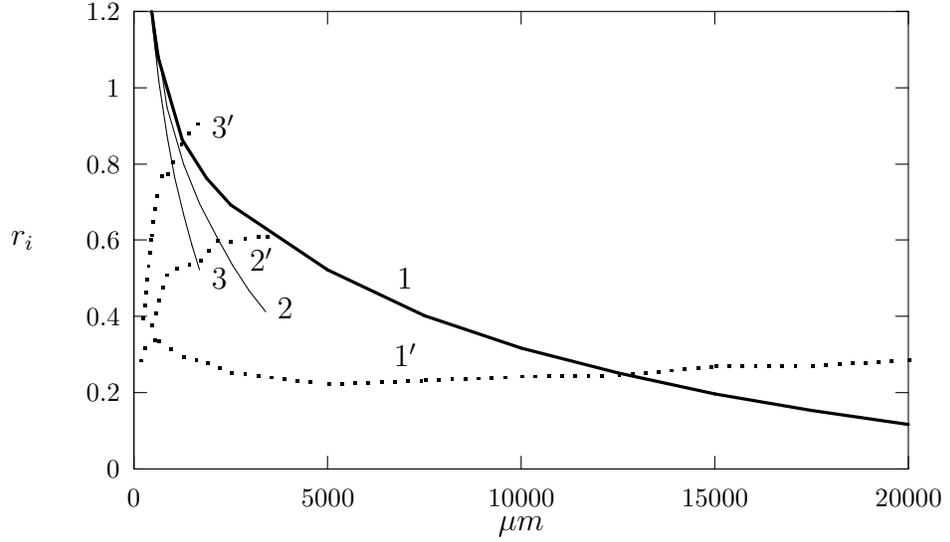
\centerline{\bf Fig.1.}
\noindent {\small Evolution of the total integral normalizied 
intensity of the
beam $r_{tot}$ and normalized incoherent part 
$r_{part}=r_{inc}/r_{tot}$
for different incidence
angles $\vartheta$. $\vartheta=0$, $r_{tot}$ (curve $1$), $r_{part}$ 
(curve $1'$); $\vartheta_F/10$ (curves $2$ and $2'$);
$\vartheta_F/5$ (curve $3$ and $3'$).
}
\end{minipage}
\end{center}
%\vskip 20pt
\par
The main result of direct simulation is that the loss of coherency
comes along with attunuation of the beam and in the transmitted
beam the coherent part prevails \cite{BobrovaJEPT1999}.
\par
Analytical results for transmission of coherent part
of X-ray can be obtained with
statistical averaging of equation (1)
using Tatarsky method (see \cite{HolyPhysStatSol1987})
as it was made in \cite{BobrovaJEPT1999} by generalization 
of the method for stratified media.
The same generalization of the method to include stratified media 
was used in the case of electron channeling in single crystals 
\cite{OgnevREffDS1993}.
The method results in additional attenuation of coherent part
of the amplitude $<A>$ due to "scattering potential" $W(x)$.

$$
%\begin{array}{l}
W (x) = (-ik/4) \int_{-\infty}^{\infty}
<\delta \varepsilon'(x,z) \delta \varepsilon'(x,z')>
dz^\prime  
$$
As it was shown in \cite{BobrovaJEPT1999} "scattering potential"
can be expessed as 

$$
W(x)\approx -{k \over 4}
%so{
{
{(\varepsilon_0-\varepsilon_1)}^2 \over{\pi {(\varepsilon_0)}^2}
}
\int_{-\infty}^{\infty}dz^\prime
\int_{-\infty}^{0} exp(-\xi^2)d\xi
\int_{0}^
{
{-R(z^\prime)\xi}\over{{(1-R^2(z^\prime))}^{1/2}}
}
exp(-{\eta}^2)d\eta
$$
$$
{\cdot} exp(-{{x^2}\over{\sigma^2}})
\eqno (2)
$$
with clear dependence on vertical coordinate $x$
where $R(z)$ is the autocorrelation coefficient,
$\sigma$ is dispersion of ${\xi}(z)$ distribution.

%\par 
%The results for coherent part of the beam obtained by averaging
%of solution of eq. (1) and integration of eq. (3) with 
%approximation (4)
%are shown of Fig.2.
The decay of coherency for particular wave modes
can be described with attenuation coefficients
$\beta_l$.
%$$
%So a
Attenuation coefficients can be found as overlap integrals
\par
\noindent
$$
\beta_l= -{k \over 2} \int{{\varphi_l}^\ast(x)
[Im(\chi(x))+W(x)]{\varphi_l}(x) }dx,
%\eqno (6)
$$
where eigenfunctions ${\varphi_j}(x)$ are solutions of equations
$$
\Delta_{\perp} {\varphi_j}(x)
=k[2k_{jz}-k Re(\chi(x))]{\varphi_j}(x).
%\eqno (5)
$$

Statistically avaraged refraction and absorption are
accounted for by normalized term
$$
{\chi}(x,z)={(<{\varepsilon} (x)> - {\varepsilon}_0)/
{\varepsilon_0}}.
$$
It can be shown for lower channeled modes that incoherent 
scattering
attenuation coefficient is proportional to $\sigma$
(see discussion above about dependence of $W(x)$ on $\sigma$)
$$
\beta_{scatter}
\sim
k^2 {(\varepsilon_0-\varepsilon_1)}^2 \sigma
\int_{-\infty}^{\infty}dz^\prime
\int_{-\infty}^{0} exp(-\xi^2/2)d\xi
\int_{0}^
{
{-R(z^\prime)\xi}\over{{(1-R^2(z^\prime))}^{1/2}}
}
exp(-{\eta/2}^2)d\eta.
$$

\par
The proportionality of losses of beam intensity to roughness 
amplitude $\sigma$
under supermall gliding angles was obtained also in the numerical 
simulation results
(\cite{BobrovaPrep1997}, Fig. 5).

\vskip 20pt
\centerline{\bf Results}

\bigskip

\par
The dependence of attenuation coefficients $\beta$ of  X-ray beam 
on the channel width $d$ between silicon plates for three lower 
modes were shown in\cite{OgnevTPL2000} and demonsrate 
$\beta \sim 1/d^3$ dependence.
Such dependence accounts for decreasing of diffractional effects 
with beam width 
$\sim \lambda /d^2$ and the effective portion of the beam that 
interacts with the surface 
$\sim \sigma /d$. When lead plates were taken into consideration 
instead of silicon
the value of attenuation coefficients became  1.5 times greater.
Increasing of $\beta$ with decreasing of energy is stronger 
than $\sim 1/E$ that 
can be accounted for by incresing of diffraction along with 
increasing of optical density of channel walls.
%%%%%%%%%%%%%%%%%%%%%%%%%%%%%

\par
Recently published experiments with Cr/C/Cr channel with length 
$L=3mm$
and width $d=1620 \AA$ of carbon layer \cite{Jark1999p9}
had shown nonmonotonous energy dependence of transmision for '0'
wave mode
(Fig.2, rombs). As it was supposed \cite{Jark1999p9} roughness
of the interfaces 
couldnot exceed $\sim 10 \AA$.
%%%%%%%%%%%%%%%%%%%%%%%%%%%%%%

%\vskip 20pt
\begin{center}
\begin{minipage}{11cm}
% GNUPLOT: LaTeX picture with emtex specials

%\input {jrk_17e2.tex}
% GNUPLOT: LaTeX picture with emtex specials
\setlength{\unitlength}{0.240900pt}
\ifx\plotpoint\undefined\newsavebox{\plotpoint}\fi
\sbox{\plotpoint}{\rule[-0.200pt]{0.400pt}{0.400pt}}%
\special{em:linewidth 0.4pt}%
\begin{picture}(1500,900)(0,0)
\font\gnuplot=cmr10 at 10pt
\gnuplot
\put(220,113){\special{em:moveto}}
\put(1436,113){\special{em:lineto}}
\put(220,113){\special{em:moveto}}
\put(240,113){\special{em:lineto}}
\put(1436,113){\special{em:moveto}}
\put(1416,113){\special{em:lineto}}
\put(198,113){\makebox(0,0)[r]{0}}
\put(220,257){\special{em:moveto}}
\put(240,257){\special{em:lineto}}
\put(1436,257){\special{em:moveto}}
\put(1416,257){\special{em:lineto}}
\put(198,257){\makebox(0,0)[r]{0.2}}
\put(220,401){\special{em:moveto}}
\put(240,401){\special{em:lineto}}
\put(1436,401){\special{em:moveto}}
\put(1416,401){\special{em:lineto}}
\put(198,401){\makebox(0,0)[r]{0.4}}
\put(220,544){\special{em:moveto}}
\put(240,544){\special{em:lineto}}
\put(1436,544){\special{em:moveto}}
\put(1416,544){\special{em:lineto}}
\put(198,544){\makebox(0,0)[r]{0.6}}
\put(220,688){\special{em:moveto}}
\put(240,688){\special{em:lineto}}
\put(1436,688){\special{em:moveto}}
\put(1416,688){\special{em:lineto}}
\put(198,688){\makebox(0,0)[r]{0.8}}
\put(220,832){\special{em:moveto}}
\put(240,832){\special{em:lineto}}
\put(1436,832){\special{em:moveto}}
\put(1416,832){\special{em:lineto}}
\put(198,832){\makebox(0,0)[r]{1}}
\put(220,113){\special{em:moveto}}
\put(220,133){\special{em:lineto}}
\put(220,832){\special{em:moveto}}
\put(220,812){\special{em:lineto}}
\put(220,68){\makebox(0,0){10}}
\put(372,113){\special{em:moveto}}
\put(372,133){\special{em:lineto}}
\put(372,832){\special{em:moveto}}
\put(372,812){\special{em:lineto}}
\put(372,68){\makebox(0,0){12}}
\put(524,113){\special{em:moveto}}
\put(524,133){\special{em:lineto}}
\put(524,832){\special{em:moveto}}
\put(524,812){\special{em:lineto}}
\put(524,68){\makebox(0,0){14}}
\put(676,113){\special{em:moveto}}
\put(676,133){\special{em:lineto}}
\put(676,832){\special{em:moveto}}
\put(676,812){\special{em:lineto}}
\put(676,68){\makebox(0,0){16}}
\put(828,113){\special{em:moveto}}
\put(828,133){\special{em:lineto}}
\put(828,832){\special{em:moveto}}
\put(828,812){\special{em:lineto}}
\put(828,68){\makebox(0,0){18}}
\put(980,113){\special{em:moveto}}
\put(980,133){\special{em:lineto}}
\put(980,832){\special{em:moveto}}
\put(980,812){\special{em:lineto}}
\put(980,68){\makebox(0,0){20}}
\put(1132,113){\special{em:moveto}}
\put(1132,133){\special{em:lineto}}
\put(1132,832){\special{em:moveto}}
\put(1132,812){\special{em:lineto}}
\put(1132,68){\makebox(0,0){22}}
\put(1284,113){\special{em:moveto}}
\put(1284,133){\special{em:lineto}}
\put(1284,832){\special{em:moveto}}
\put(1284,812){\special{em:lineto}}
\put(1284,68){\makebox(0,0){24}}
\put(1436,113){\special{em:moveto}}
\put(1436,133){\special{em:lineto}}
\put(1436,832){\special{em:moveto}}
\put(1436,812){\special{em:lineto}}
\put(1436,68){\makebox(0,0){26}}
\put(220,113){\special{em:moveto}}
\put(1436,113){\special{em:lineto}}
\put(1436,832){\special{em:lineto}}
\put(220,832){\special{em:lineto}}
\put(220,113){\special{em:lineto}}
\put(45,472){\makebox(0,0){$T$}}
\put(828,23){\makebox(0,0){$E,~keV$}}
%\put(828,877){\makebox(0,0){Fig.1}}
\put(1284,580){\makebox(0,0){$1$}}
\put(904,760){\makebox(0,0){$2$}}
\put(1132,688){\makebox(0,0){$3$}}
\sbox{\plotpoint}{\rule[-0.500pt]{1.000pt}{1.000pt}}%
\special{em:linewidth 1.0pt}%
\put(220,228){\usebox{\plotpoint}}
\multiput(220,228)(14.146,15.188){14}{\usebox{\plotpoint}}
\multiput(410,432)(17.130,11.720){11}{\usebox{\plotpoint}}
\multiput(600,562)(11.443,-17.316){13}{\usebox{\plotpoint}}
\multiput(752,332)(14.264,15.077){16}{\usebox{\plotpoint}}
\multiput(980,573)(20.618,-2.387){19}{\usebox{\plotpoint}}
\put(1360,529){\usebox{\plotpoint}}
\sbox{\plotpoint}{\rule[-0.200pt]{0.400pt}{0.400pt}}%
\special{em:linewidth 0.4pt}%
\put(220,235){\special{em:moveto}}
\put(410,432){\special{em:lineto}}
\put(600,578){\special{em:lineto}}
\put(752,651){\special{em:lineto}}
\put(980,722){\special{em:lineto}}
\put(1360,773){\special{em:lineto}}
\sbox{\plotpoint}{\rule[-0.400pt]{0.800pt}{0.800pt}}%
\special{em:linewidth 0.8pt}%
\put(220,274){\special{em:moveto}}
\put(410,448){\special{em:lineto}}
\put(600,565){\special{em:lineto}}
\put(752,642){\special{em:lineto}}
\put(980,708){\special{em:lineto}}
\put(1360,772){\special{em:lineto}}
\sbox{\plotpoint}{\rule[-0.500pt]{1.000pt}{1.000pt}}%
\special{em:linewidth 1.0pt}%
\put(296,199){\raisebox{-.8pt}{\makebox(0,0){$\Diamond$}}}
\put(410,401){\raisebox{-.8pt}{\makebox(0,0){$\Diamond$}}}
\put(600,598){\raisebox{-.8pt}{\makebox(0,0){$\Diamond$}}}
\put(752,437){\raisebox{-.8pt}{\makebox(0,0){$\Diamond$}}}
\put(980,688){\raisebox{-.8pt}{\makebox(0,0){$\Diamond$}}}
\put(1360,667){\raisebox{-.8pt}{\makebox(0,0){$\Diamond$}}}
\end{picture}

%\vskip 15pt
\par
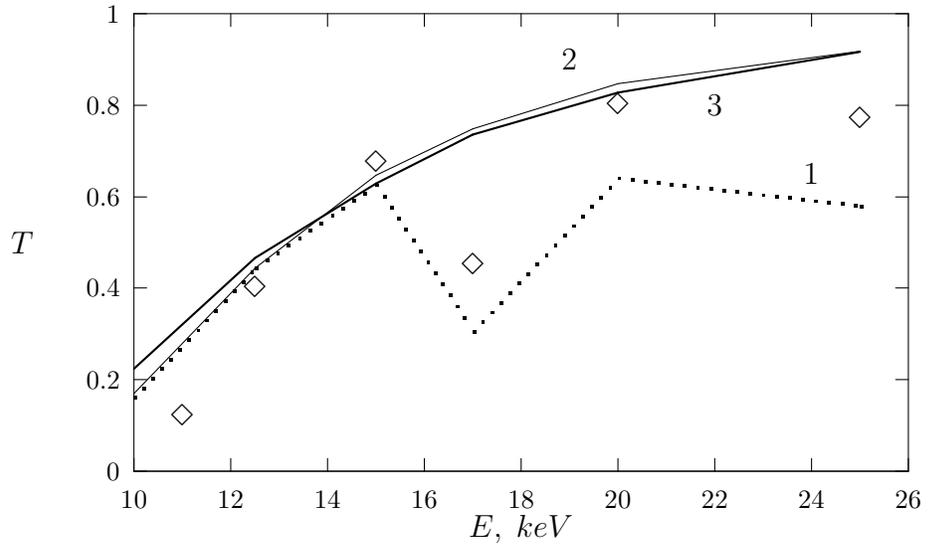
\centerline{\bf Fig.2.}
\noindent {\small 
Calculated dependence of basic '0' wave mode transmission $T$
in Cr/C/Cr channel on X-ray energy. $L=3mm$, $d=1620 \AA$.
Deformation amplitude $a=120 \AA$, 
period $\Lambda=100 \mu m$ 
(curve 1), $500\mu m$ (2), $1000\mu m$ (3). 
$\sigma=0 \AA$.
Experimental points of W.~Jark et al [2] are shown by 
rombs.
%asterisks.
}
\end{minipage}
\end{center}
%\vskip 20pt

\par
Direct numerical simulation of the transmission of X-ray beam 
with equation (1) was 
developed to investigate the dependence of '0' and '1' modes 
transmission on roughness
amplitude. The account for roughness decreases transmission 
of the basic mode with
$E=17keV$ by 1.3~\% for $\sigma=10 \AA$ and by 
5~\% for $\sigma=20 \AA$ (see Fig.3
) 
that cannot explain prominent depression of experimental results 
on Fig.2.
%%%
%%%
 
%\vskip 20pt
\begin{center}
\begin{minipage}{11cm}
% GNUPLOT: LaTeX picture with emtex specials

%\input {twomod17.tex}
% GNUPLOT: LaTeX picture with emtex specials
\setlength{\unitlength}{0.240900pt}
\ifx\plotpoint\undefined\newsavebox{\plotpoint}\fi
\sbox{\plotpoint}{\rule[-0.200pt]{0.400pt}{0.400pt}}%
\special{em:linewidth 0.4pt}%
\begin{picture}(1500,900)(0,0)
\font\gnuplot=cmr10 at 10pt
\gnuplot
\put(220,113){\special{em:moveto}}
\put(1436,113){\special{em:lineto}}
\put(220,113){\special{em:moveto}}
\put(220,877){\special{em:lineto}}
\put(220,113){\special{em:moveto}}
\put(240,113){\special{em:lineto}}
\put(1436,113){\special{em:moveto}}
\put(1416,113){\special{em:lineto}}
\put(198,113){\makebox(0,0)[r]{0}}
\put(220,266){\special{em:moveto}}
\put(240,266){\special{em:lineto}}
\put(1436,266){\special{em:moveto}}
\put(1416,266){\special{em:lineto}}
\put(198,266){\makebox(0,0)[r]{0.2}}
\put(220,419){\special{em:moveto}}
\put(240,419){\special{em:lineto}}
\put(1436,419){\special{em:moveto}}
\put(1416,419){\special{em:lineto}}
\put(198,419){\makebox(0,0)[r]{0.4}}
\put(220,571){\special{em:moveto}}
\put(240,571){\special{em:lineto}}
\put(1436,571){\special{em:moveto}}
\put(1416,571){\special{em:lineto}}
\put(198,571){\makebox(0,0)[r]{0.6}}
\put(220,724){\special{em:moveto}}
\put(240,724){\special{em:lineto}}
\put(1436,724){\special{em:moveto}}
\put(1416,724){\special{em:lineto}}
\put(198,724){\makebox(0,0)[r]{0.8}}
\put(220,877){\special{em:moveto}}
\put(240,877){\special{em:lineto}}
\put(1436,877){\special{em:moveto}}
\put(1416,877){\special{em:lineto}}
\put(198,877){\makebox(0,0)[r]{1}}
\put(220,113){\special{em:moveto}}
\put(220,133){\special{em:lineto}}
\put(220,877){\special{em:moveto}}
\put(220,857){\special{em:lineto}}
\put(220,68){\makebox(0,0){0}}
\put(524,113){\special{em:moveto}}
\put(524,133){\special{em:lineto}}
\put(524,877){\special{em:moveto}}
\put(524,857){\special{em:lineto}}
\put(524,68){\makebox(0,0){5}}
\put(828,113){\special{em:moveto}}
\put(828,133){\special{em:lineto}}
\put(828,877){\special{em:moveto}}
\put(828,857){\special{em:lineto}}
\put(828,68){\makebox(0,0){10}}
\put(1132,113){\special{em:moveto}}
\put(1132,133){\special{em:lineto}}
\put(1132,877){\special{em:moveto}}
\put(1132,857){\special{em:lineto}}
\put(1132,68){\makebox(0,0){15}}
\put(1436,113){\special{em:moveto}}
\put(1436,133){\special{em:lineto}}
\put(1436,877){\special{em:moveto}}
\put(1436,857){\special{em:lineto}}
\put(1436,68){\makebox(0,0){20}}
\put(220,113){\special{em:moveto}}
\put(1436,113){\special{em:lineto}}
\put(1436,877){\special{em:lineto}}
\put(220,877){\special{em:lineto}}
\put(220,113){\special{em:lineto}}
\put(45,495){\makebox(0,0){$T$}}
\put(828,23){\makebox(0,0){roughness $\sigma, \AA$}}
\put(1132,724){\makebox(0,0){$'0'$}}
\put(1314,610){\makebox(0,0){$'1'$}}
\put(220,701){\special{em:moveto}}
\put(828,694){\special{em:lineto}}
\put(1436,671){\special{em:lineto}}
\sbox{\plotpoint}{\rule[-0.500pt]{1.000pt}{1.000pt}}%
\special{em:linewidth 1.0pt}%
\put(220,655){\usebox{\plotpoint}}
\multiput(220,655)(20.730,-1.023){30}{\usebox{\plotpoint}}
\multiput(828,625)(20.486,-3.336){30}{\usebox{\plotpoint}}
\end{picture}

%\end{picture}

%\vskip 15pt

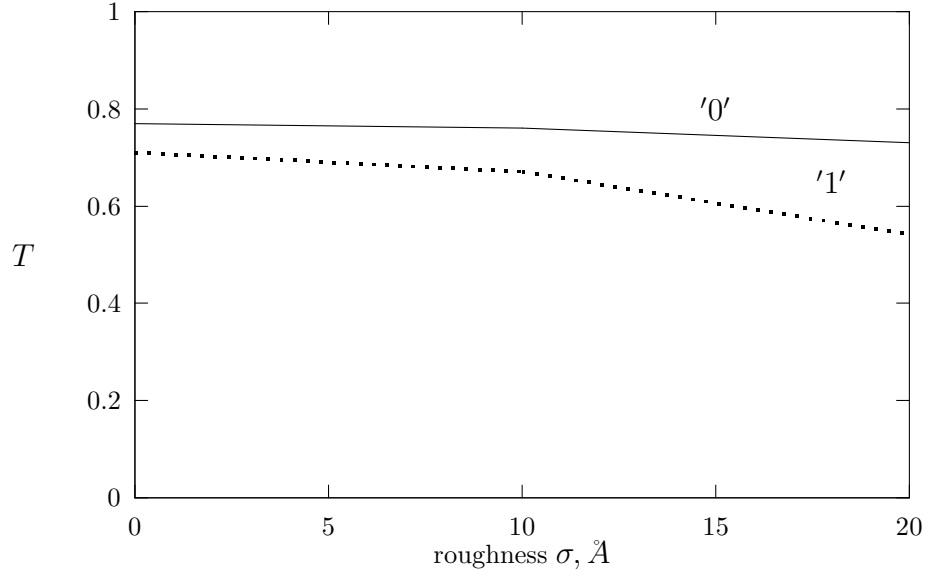
\centerline{\bf Fig.3}
\noindent {\small Dependence of transmission
of 17 keV X-ray beam in the channel Cr/C/Cr
width C layer $d=1620 \AA$ for modes '0' and '1' on roughness 
$\sigma$, $z_{corr} = 5\mu m$. 
%Attenuation due to scattering $\beta_{scatter}$
%is shown by dashed lines, attenuation due to absorption
%$\beta_{absorp}$ is shown by solid lines.
}
\end{minipage}
\end{center}
%\vskip 20pt

%%%
%%%
\par
For the expanation of anomalous dependence of 17keV radiation 
basic mode transmision
through Cr/C/Cr channels periodic deformation of the layers were 
taken into acount.
The results are shown on Fig.2 for deformation amplitude 
$a=120 \AA$ and periods
$\Lambda=100 \mu m$ (curve 1), $500 \mu m$ (2) and 
$1000 \mu m$ (3).
\par
The dependence of transmission on deformation period $\Lambda$ 
for $E=17keV$ $a=120 \AA$ 
and without roughness (the effect of roughness was not important; 
see Fig.3 above) 
is shown on Fig.4. Several resonanses can be recognised in short 
$\Lambda$ region.
So the results shown on Fig.2 and Fig.4 are similar to the 
complicated effects of strong wave function transformation 
of channeled electrons in superlattices 
\cite{BobrovaPhysStatSol1997}.
%%%%%%%%%%%%%%%%%%%%%%%%%%%%
\par 
Thus the depression of the transmission for $E=17keV$ on Fig.2 
observed in \cite{Jark1999p9} 
can be result of the periodic corrugation of Cr/C interface and 
wave mode interference.

\par
To clear out the mechanism of decay of x-ray beam in thin film 
waveguide with periodic
perpurbations both decay of total intensity and basic mode "0" 
intensity on the distance 
was investigated for different periods $\Lambda$.
\par
In the case of small scale perturbations 
($\Lambda \le 45 \mu m$) basic mode intensity
decreases nearly the same as the whole beam. And in the case 
of resonant perturbation
$\Lambda=45\mu m$ the intensity of basic mode  is subjected 
to strong oscillations
with the period $\Lambda /2$, decreasing at the distance 
$z=3000\AA$ to 0.03 part 
of the initial value. In the nonresonance case $\Lambda=40\mu m$ 
the basic mode "0" 
oscilations are substantial only near the entrance to the carbon 
channel. Intensity 
at the distance $z=3000\AA$ on exit of the channel decreases 
to 0.6 of the initial value. 
\par
For the period ($\Lambda =1000 \mu m$) the dependence of total 
intensity and basic mode 
"0" intensity are shown on Fig.5 with curves $2$ (points) and $2'$ 
(solid). Curves $1$ and $1'$ correspond  to the direct channel. 
Pulsations on the curve $1'$ are due to calcuiation uncertainties.
From the Fig.5 it is seen that in the case of large scale 
perturbations the decreasing
of total intensity slightly differes from the streight channel. 
But decreasing of basic mode
having the oscilation manner with the period $\Lambda /2$, 
may reach nearly 0.1 of the
initial value. That is why the influence of large scale perturbations 
must result in substantial
increasing of angular spread of the beam at the exit of the channel.

\vskip 20pt
\centerline{\bf Discussion}
\bigskip

The investigations developed show strong influence of deformations 
of the channel on the transmission of x-ray channeled beams. 
Small scale random perturbations of the surface with the roughness
amplitude up to $20\AA$
do not effect considerably the transmission of X-rays in comparison
with diffraction effects
that determine the decay of intensity in the channel in the case.
The X-ray transmission is the most sensitive to the resonant 
periodical  perturbations 
of the channel corresponding to the pendulum oscillations of modes 
"0" and "1".
In this case nearly complete dempening of the beam due to transfer 
of basic mode "0"
to upper modes which decay 
rapidly\cite{BobrovaJEPT1999,OgnevTPL2000}.
In the case of large periods of deformations of the channel effective 
transfer of the beam to the higher
modes takes place but it do not succed in substancial change of 
total intensity.
It is worth noting that the effect of abnormal energy dependence 
of transmission of the beam
through Cr/C/Cr channel that was observed in \cite{Jark1999p9} 
dissapeared after 
the technology of production of X-ray waveguides was improved 
\cite{Jark_priv}
that can serve as the confirmation of the results of the present 
work.
\par
The results of the present work can be used for creation of new 
type of tunable 
X-ray filters for formation of thin beams of synchrotron X-ray 
radiation.

\newpage

%\newpage

%\vskip 20pt
\begin{center}
\begin{minipage}{11cm}
% GNUPLOT: LaTeX picture with emtex specials

%\input {f1720kev.tex}
% GNUPLOT: LaTeX picture with emtex specials
\setlength{\unitlength}{0.240900pt}
\ifx\plotpoint\undefined\newsavebox{\plotpoint}\fi
\sbox{\plotpoint}{\rule[-0.200pt]{0.400pt}{0.400pt}}%
\special{em:linewidth 0.4pt}%
\begin{picture}(1500,900)(0,0)
\font\gnuplot=cmr10 at 10pt
\gnuplot
\put(220,113){\special{em:moveto}}
\put(1436,113){\special{em:lineto}}
\put(220,113){\special{em:moveto}}
\put(220,832){\special{em:lineto}}
\put(220,113){\special{em:moveto}}
\put(240,113){\special{em:lineto}}
\put(1436,113){\special{em:moveto}}
\put(1416,113){\special{em:lineto}}
\put(198,113){\makebox(0,0)[r]{0}}
\put(220,257){\special{em:moveto}}
\put(240,257){\special{em:lineto}}
\put(1436,257){\special{em:moveto}}
\put(1416,257){\special{em:lineto}}
\put(198,257){\makebox(0,0)[r]{0.2}}
\put(220,401){\special{em:moveto}}
\put(240,401){\special{em:lineto}}
\put(1436,401){\special{em:moveto}}
\put(1416,401){\special{em:lineto}}
\put(198,401){\makebox(0,0)[r]{0.4}}
\put(220,544){\special{em:moveto}}
\put(240,544){\special{em:lineto}}
\put(1436,544){\special{em:moveto}}
\put(1416,544){\special{em:lineto}}
\put(198,544){\makebox(0,0)[r]{0.6}}
\put(220,688){\special{em:moveto}}
\put(240,688){\special{em:lineto}}
\put(1436,688){\special{em:moveto}}
\put(1416,688){\special{em:lineto}}
\put(198,688){\makebox(0,0)[r]{0.8}}
\put(220,832){\special{em:moveto}}
\put(240,832){\special{em:lineto}}
\put(1436,832){\special{em:moveto}}
\put(1416,832){\special{em:lineto}}
\put(198,832){\makebox(0,0)[r]{1}}
\put(220,113){\special{em:moveto}}
\put(220,133){\special{em:lineto}}
\put(220,832){\special{em:moveto}}
\put(220,812){\special{em:lineto}}
\put(220,68){\makebox(0,0){0}}
\put(342,113){\special{em:moveto}}
\put(342,133){\special{em:lineto}}
\put(342,832){\special{em:moveto}}
\put(342,812){\special{em:lineto}}
\put(342,68){\makebox(0,0){50}}
\put(463,113){\special{em:moveto}}
\put(463,133){\special{em:lineto}}
\put(463,832){\special{em:moveto}}
\put(463,812){\special{em:lineto}}
\put(463,68){\makebox(0,0){100}}
\put(585,113){\special{em:moveto}}
\put(585,133){\special{em:lineto}}
\put(585,832){\special{em:moveto}}
\put(585,812){\special{em:lineto}}
\put(585,68){\makebox(0,0){150}}
\put(706,113){\special{em:moveto}}
\put(706,133){\special{em:lineto}}
\put(706,832){\special{em:moveto}}
\put(706,812){\special{em:lineto}}
\put(706,68){\makebox(0,0){200}}
\put(828,113){\special{em:moveto}}
\put(828,133){\special{em:lineto}}
\put(828,832){\special{em:moveto}}
\put(828,812){\special{em:lineto}}
\put(828,68){\makebox(0,0){250}}
\put(950,113){\special{em:moveto}}
\put(950,133){\special{em:lineto}}
\put(950,832){\special{em:moveto}}
\put(950,812){\special{em:lineto}}
\put(950,68){\makebox(0,0){300}}
\put(1071,113){\special{em:moveto}}
\put(1071,133){\special{em:lineto}}
\put(1071,832){\special{em:moveto}}
\put(1071,812){\special{em:lineto}}
\put(1071,68){\makebox(0,0){350}}
\put(1193,113){\special{em:moveto}}
\put(1193,133){\special{em:lineto}}
\put(1193,832){\special{em:moveto}}
\put(1193,812){\special{em:lineto}}
\put(1193,68){\makebox(0,0){400}}
\put(1314,113){\special{em:moveto}}
\put(1314,133){\special{em:lineto}}
\put(1314,832){\special{em:moveto}}
\put(1314,812){\special{em:lineto}}
\put(1314,68){\makebox(0,0){450}}
\put(1436,113){\special{em:moveto}}
\put(1436,133){\special{em:lineto}}
\put(1436,832){\special{em:moveto}}
\put(1436,812){\special{em:lineto}}
\put(1436,68){\makebox(0,0){500}}
\put(220,113){\special{em:moveto}}
\put(1436,113){\special{em:lineto}}
\put(1436,832){\special{em:lineto}}
\put(220,832){\special{em:lineto}}
\put(220,113){\special{em:lineto}}
\put(45,472){\makebox(0,0){$T$}}
\put(828,23){\makebox(0,0){$\Lambda, \mu m$}}
%\put(828,877){\makebox(0,0){Fig.2}}
\put(269,473){\special{em:moveto}}
\put(281,578){\special{em:lineto}}
\put(293,424){\special{em:lineto}}
\put(305,366){\special{em:lineto}}
\put(317,558){\special{em:lineto}}
\put(329,135){\special{em:lineto}}
\put(342,260){\special{em:lineto}}
\put(354,465){\special{em:lineto}}
\put(366,437){\special{em:lineto}}
\put(378,471){\special{em:lineto}}
\put(390,406){\special{em:lineto}}
\put(402,171){\special{em:lineto}}
\put(415,500){\special{em:lineto}}
\put(427,507){\special{em:lineto}}
\put(439,155){\special{em:lineto}}
\put(451,154){\special{em:lineto}}
\put(463,332){\special{em:lineto}}
\put(475,574){\special{em:lineto}}
\put(488,627){\special{em:lineto}}
\put(500,631){\special{em:lineto}}
\put(512,623){\special{em:lineto}}
\put(524,626){\special{em:lineto}}
\put(536,634){\special{em:lineto}}
\put(548,634){\special{em:lineto}}
\put(560,630){\special{em:lineto}}
\put(573,629){\special{em:lineto}}
\put(585,624){\special{em:lineto}}
\put(597,617){\special{em:lineto}}
\put(609,606){\special{em:lineto}}
\put(621,599){\special{em:lineto}}
\put(633,593){\special{em:lineto}}
\put(646,579){\special{em:lineto}}
\put(658,566){\special{em:lineto}}
\put(670,520){\special{em:lineto}}
\put(682,490){\special{em:lineto}}
\put(694,375){\special{em:lineto}}
\put(706,319){\special{em:lineto}}
\put(719,367){\special{em:lineto}}
\put(731,372){\special{em:lineto}}
\put(743,375){\special{em:lineto}}
\put(755,390){\special{em:lineto}}
\put(767,440){\special{em:lineto}}
\put(779,342){\special{em:lineto}}
\put(792,510){\special{em:lineto}}
\put(804,408){\special{em:lineto}}
\put(816,452){\special{em:lineto}}
\put(828,581){\special{em:lineto}}
\put(840,532){\special{em:lineto}}
\put(852,459){\special{em:lineto}}
\put(864,609){\special{em:lineto}}
\put(877,626){\special{em:lineto}}
\put(889,580){\special{em:lineto}}
\put(901,526){\special{em:lineto}}
\put(913,609){\special{em:lineto}}
\put(925,642){\special{em:lineto}}
\put(937,645){\special{em:lineto}}
\put(950,627){\special{em:lineto}}
\put(962,597){\special{em:lineto}}
\put(974,613){\special{em:lineto}}
\put(986,643){\special{em:lineto}}
\put(998,650){\special{em:lineto}}
\put(1010,651){\special{em:lineto}}
\put(1023,651){\special{em:lineto}}
\put(1035,644){\special{em:lineto}}
\put(1047,639){\special{em:lineto}}
\put(1059,641){\special{em:lineto}}
\put(1071,649){\special{em:lineto}}
\put(1083,651){\special{em:lineto}}
\put(1096,652){\special{em:lineto}}
\put(1108,654){\special{em:lineto}}
\put(1120,656){\special{em:lineto}}
\put(1132,656){\special{em:lineto}}
\put(1144,656){\special{em:lineto}}
\put(1156,656){\special{em:lineto}}
\put(1168,656){\special{em:lineto}}
\put(1181,653){\special{em:lineto}}
\put(1193,651){\special{em:lineto}}
\put(1205,650){\special{em:lineto}}
\put(1217,651){\special{em:lineto}}
\put(1229,653){\special{em:lineto}}
\put(1241,655){\special{em:lineto}}
\put(1254,655){\special{em:lineto}}
\put(1266,655){\special{em:lineto}}
\put(1278,654){\special{em:lineto}}
\put(1290,654){\special{em:lineto}}
\put(1302,653){\special{em:lineto}}
\put(1314,652){\special{em:lineto}}
\put(1327,650){\special{em:lineto}}
\put(1339,649){\special{em:lineto}}
\put(1351,649){\special{em:lineto}}
\put(1363,649){\special{em:lineto}}
\put(1375,649){\special{em:lineto}}
\put(1387,650){\special{em:lineto}}
\put(1400,651){\special{em:lineto}}
\put(1412,651){\special{em:lineto}}
\put(1424,651){\special{em:lineto}}
\put(1436,651){\special{em:lineto}}
\sbox{\plotpoint}{\rule[-0.500pt]{1.000pt}{1.000pt}}%
\special{em:linewidth 1.0pt}%
\put(269,682){\usebox{\plotpoint}}
\multiput(269,682)(2.449,-20.611){5}{\usebox{\plotpoint}}
\multiput(281,581)(5.964,19.880){2}{\usebox{\plotpoint}}
\multiput(293,621)(1.151,-20.724){11}{\usebox{\plotpoint}}
\multiput(305,405)(6.250,19.792){2}{\usebox{\plotpoint}}
\multiput(317,443)(2.192,20.639){5}{\usebox{\plotpoint}}
\multiput(329,556)(1.289,-20.715){10}{\usebox{\plotpoint}}
\put(344.65,349.65){\usebox{\plotpoint}}
\multiput(354,359)(2.015,20.657){6}{\usebox{\plotpoint}}
\multiput(366,482)(3.196,20.508){4}{\usebox{\plotpoint}}
\put(384.28,546.97){\usebox{\plotpoint}}
\put(396.98,540.65){\usebox{\plotpoint}}
\multiput(402,544)(2.352,-20.622){5}{\usebox{\plotpoint}}
\put(417.55,432.76){\usebox{\plotpoint}}
\multiput(427,443)(2.100,-20.649){6}{\usebox{\plotpoint}}
\multiput(439,325)(0.915,20.735){13}{\usebox{\plotpoint}}
\put(455.61,587.78){\usebox{\plotpoint}}
\multiput(463,573)(0.736,-20.742){17}{\usebox{\plotpoint}}
\put(481.41,252.74){\usebox{\plotpoint}}
\multiput(488,271)(4.006,-20.365){3}{\usebox{\plotpoint}}
\multiput(500,210)(0.783,20.741){16}{\usebox{\plotpoint}}
\multiput(512,528)(1.736,20.683){7}{\usebox{\plotpoint}}
\put(531.16,686.52){\usebox{\plotpoint}}
\put(545.08,698.51){\usebox{\plotpoint}}
\multiput(548,699)(20.473,-3.412){0}{\usebox{\plotpoint}}
\multiput(560,697)(9.004,-18.701){2}{\usebox{\plotpoint}}
\put(579.77,685.79){\usebox{\plotpoint}}
\put(592.44,698.62){\usebox{\plotpoint}}
\multiput(597,699)(20.473,-3.412){0}{\usebox{\plotpoint}}
\put(613.02,697.00){\usebox{\plotpoint}}
\multiput(621,697)(20.473,-3.412){0}{\usebox{\plotpoint}}
\put(633.55,694.75){\usebox{\plotpoint}}
\put(653.04,689.00){\usebox{\plotpoint}}
\multiput(658,689)(19.690,-6.563){0}{\usebox{\plotpoint}}
\put(672.05,682.61){\usebox{\plotpoint}}
\put(687.19,669.27){\usebox{\plotpoint}}
\multiput(694,667)(19.690,-6.563){0}{\usebox{\plotpoint}}
\put(706.79,662.52){\usebox{\plotpoint}}
\put(723.54,650.46){\usebox{\plotpoint}}
\put(737.92,635.50){\usebox{\plotpoint}}
\put(752.38,620.62){\usebox{\plotpoint}}
\multiput(755,618)(3.079,-20.526){4}{\usebox{\plotpoint}}
\multiput(767,538)(2.743,-20.573){4}{\usebox{\plotpoint}}
\multiput(779,448)(3.738,-20.416){3}{\usebox{\plotpoint}}
\multiput(792,377)(5.579,19.992){3}{\usebox{\plotpoint}}
\put(809.90,438.18){\usebox{\plotpoint}}
\put(816.92,457.31){\usebox{\plotpoint}}
\multiput(828,461)(5.461,-20.024){2}{\usebox{\plotpoint}}
\multiput(840,417)(2.970,20.542){4}{\usebox{\plotpoint}}
\put(854.21,503.50){\usebox{\plotpoint}}
\multiput(864,519)(2.480,-20.607){6}{\usebox{\plotpoint}}
\multiput(877,411)(1.445,20.705){8}{\usebox{\plotpoint}}
\put(894.46,572.54){\usebox{\plotpoint}}
\multiput(901,560)(2.335,-20.624){5}{\usebox{\plotpoint}}
\multiput(913,454)(1.689,20.687){7}{\usebox{\plotpoint}}
\multiput(925,601)(3.459,20.465){4}{\usebox{\plotpoint}}
\multiput(937,672)(5.426,-20.034){2}{\usebox{\plotpoint}}
\multiput(950,624)(4.276,-20.310){3}{\usebox{\plotpoint}}
\multiput(962,567)(3.117,20.520){4}{\usebox{\plotpoint}}
\multiput(974,646)(5.034,20.136){2}{\usebox{\plotpoint}}
\put(989.30,692.90){\usebox{\plotpoint}}
\multiput(998,690)(5.348,-20.055){2}{\usebox{\plotpoint}}
\put(1015.55,641.58){\usebox{\plotpoint}}
\multiput(1023,637)(6.250,19.792){2}{\usebox{\plotpoint}}
\put(1040.08,687.70){\usebox{\plotpoint}}
\put(1049.01,705.67){\usebox{\plotpoint}}
\put(1068.70,705.77){\usebox{\plotpoint}}
\put(1082.93,691.08){\usebox{\plotpoint}}
\multiput(1083,691)(18.275,-9.840){0}{\usebox{\plotpoint}}
\put(1100.52,687.77){\usebox{\plotpoint}}
\put(1115.17,702.37){\usebox{\plotpoint}}
\put(1131.10,715.40){\usebox{\plotpoint}}
\multiput(1132,716)(20.473,3.412){0}{\usebox{\plotpoint}}
\put(1151.50,718.00){\usebox{\plotpoint}}
\multiput(1156,718)(20.473,-3.412){0}{\usebox{\plotpoint}}
\put(1172.08,715.69){\usebox{\plotpoint}}
\put(1192.65,713.06){\usebox{\plotpoint}}
\end{picture}

%\vskip 15pt

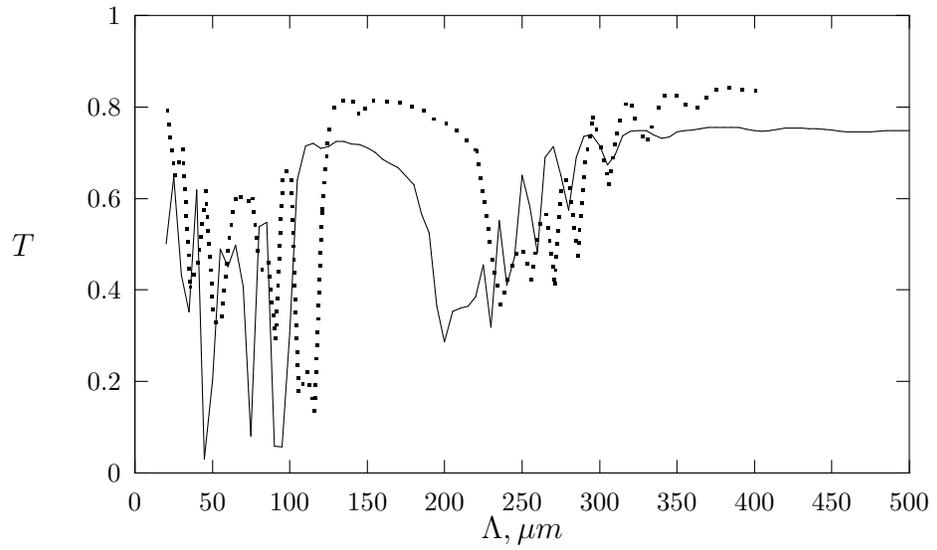
\centerline{\bf Fig.~4}
\noindent {\small 
The dependence of basic '0' wave mode transmission $T$
of $E=17keV$ X-ray beam in Cr/C/Cr $d=1620 \AA$ channel 
on the deformation period $\Lambda$.
$\sigma = 0 \AA$, $L=3mm$,  $a= 120 \AA$.
}
\end{minipage}
\end{center}
\newpage
\vskip 20pt
\begin{center}
\begin{minipage}{11cm}
% GNUPLOT: LaTeX picture with emtex specials

%\input {r123_2nn.tex}
% GNUPLOT: LaTeX picture with emtex specials
\setlength{\unitlength}{0.240900pt}
\ifx\plotpoint\undefined\newsavebox{\plotpoint}\fi
\sbox{\plotpoint}{\rule[-0.200pt]{0.400pt}{0.400pt}}%
\special{em:linewidth 0.4pt}%
\begin{picture}(1500,900)(0,0)
\font\gnuplot=cmr10 at 10pt
\gnuplot
\put(220,113){\special{em:moveto}}
\put(1436,113){\special{em:lineto}}
\put(220,113){\special{em:moveto}}
\put(220,877){\special{em:lineto}}
\put(220,113){\special{em:moveto}}
\put(240,113){\special{em:lineto}}
\put(1436,113){\special{em:moveto}}
\put(1416,113){\special{em:lineto}}
\put(198,113){\makebox(0,0)[r]{0}}
\put(220,252){\special{em:moveto}}
\put(240,252){\special{em:lineto}}
\put(1436,252){\special{em:moveto}}
\put(1416,252){\special{em:lineto}}
\put(198,252){\makebox(0,0)[r]{0.2}}
\put(220,391){\special{em:moveto}}
\put(240,391){\special{em:lineto}}
\put(1436,391){\special{em:moveto}}
\put(1416,391){\special{em:lineto}}
\put(198,391){\makebox(0,0)[r]{0.4}}
\put(220,530){\special{em:moveto}}
\put(240,530){\special{em:lineto}}
\put(1436,530){\special{em:moveto}}
\put(1416,530){\special{em:lineto}}
\put(198,530){\makebox(0,0)[r]{0.6}}
\put(220,669){\special{em:moveto}}
\put(240,669){\special{em:lineto}}
\put(1436,669){\special{em:moveto}}
\put(1416,669){\special{em:lineto}}
\put(198,669){\makebox(0,0)[r]{0.8}}
\put(220,808){\special{em:moveto}}
\put(240,808){\special{em:lineto}}
\put(1436,808){\special{em:moveto}}
\put(1416,808){\special{em:lineto}}
\put(198,808){\makebox(0,0)[r]{1}}
\put(220,113){\special{em:moveto}}
\put(220,133){\special{em:lineto}}
\put(220,877){\special{em:moveto}}
\put(220,857){\special{em:lineto}}
\put(220,68){\makebox(0,0){0}}
\put(423,113){\special{em:moveto}}
\put(423,133){\special{em:lineto}}
\put(423,877){\special{em:moveto}}
\put(423,857){\special{em:lineto}}
\put(423,68){\makebox(0,0){500}}
\put(625,113){\special{em:moveto}}
\put(625,133){\special{em:lineto}}
\put(625,877){\special{em:moveto}}
\put(625,857){\special{em:lineto}}
\put(625,68){\makebox(0,0){1000}}
\put(828,113){\special{em:moveto}}
\put(828,133){\special{em:lineto}}
\put(828,877){\special{em:moveto}}
\put(828,857){\special{em:lineto}}
\put(828,68){\makebox(0,0){1500}}
\put(1031,113){\special{em:moveto}}
\put(1031,133){\special{em:lineto}}
\put(1031,877){\special{em:moveto}}
\put(1031,857){\special{em:lineto}}
\put(1031,68){\makebox(0,0){2000}}
\put(1233,113){\special{em:moveto}}
\put(1233,133){\special{em:lineto}}
\put(1233,877){\special{em:moveto}}
\put(1233,857){\special{em:lineto}}
\put(1233,68){\makebox(0,0){2500}}
\put(1436,113){\special{em:moveto}}
\put(1436,133){\special{em:lineto}}
\put(1436,877){\special{em:moveto}}
\put(1436,857){\special{em:lineto}}
\put(1436,68){\makebox(0,0){3000}}
\put(220,113){\special{em:moveto}}
\put(1436,113){\special{em:lineto}}
\put(1436,877){\special{em:lineto}}
\put(220,877){\special{em:lineto}}
\put(220,113){\special{em:lineto}}
\put(45,495){\makebox(0,0){$T$}}
\put(828,23){\makebox(0,0){$z, \mu m$}}
\put(1031,738){\makebox(0,0){$1'$}}
\put(1112,738){\makebox(0,0){$1$}}
\put(909,509){\makebox(0,0){$2'$}}
\put(950,599){\makebox(0,0){$2$}}
\put(1233,460){\makebox(0,0){$\Lambda=1000 \mu m$}}
\put(1233,599){\makebox(0,0){$\uparrow$}}
\put(1233,321){\makebox(0,0){$\downarrow$}}
\put(958,721){\makebox(0,0){$\swarrow$}}
\put(1010,662){\makebox(0,0){$\nearrow$}}
\put(233,803){\special{em:moveto}}
\put(246,796){\special{em:lineto}}
\put(259,792){\special{em:lineto}}
\put(272,791){\special{em:lineto}}
\put(285,785){\special{em:lineto}}
\put(298,781){\special{em:lineto}}
\put(311,778){\special{em:lineto}}
\put(324,782){\special{em:lineto}}
\put(337,782){\special{em:lineto}}
\put(350,781){\special{em:lineto}}
\put(363,782){\special{em:lineto}}
\put(376,784){\special{em:lineto}}
\put(389,783){\special{em:lineto}}
\put(402,776){\special{em:lineto}}
\put(415,772){\special{em:lineto}}
\put(428,768){\special{em:lineto}}
\put(441,766){\special{em:lineto}}
\put(453,759){\special{em:lineto}}
\put(466,758){\special{em:lineto}}
\put(479,757){\special{em:lineto}}
\put(492,760){\special{em:lineto}}
\put(505,758){\special{em:lineto}}
\put(518,759){\special{em:lineto}}
\put(531,760){\special{em:lineto}}
\put(544,762){\special{em:lineto}}
\put(557,756){\special{em:lineto}}
\put(570,751){\special{em:lineto}}
\put(583,748){\special{em:lineto}}
\put(596,746){\special{em:lineto}}
\put(609,741){\special{em:lineto}}
\put(622,736){\special{em:lineto}}
\put(635,735){\special{em:lineto}}
\put(648,737){\special{em:lineto}}
\put(661,738){\special{em:lineto}}
\put(674,736){\special{em:lineto}}
\put(687,737){\special{em:lineto}}
\put(700,738){\special{em:lineto}}
\put(713,739){\special{em:lineto}}
\put(726,732){\special{em:lineto}}
\put(739,728){\special{em:lineto}}
\put(752,725){\special{em:lineto}}
\put(765,724){\special{em:lineto}}
\put(778,717){\special{em:lineto}}
\put(791,714){\special{em:lineto}}
\put(804,714){\special{em:lineto}}
\put(817,717){\special{em:lineto}}
\put(830,716){\special{em:lineto}}
\put(843,716){\special{em:lineto}}
\put(856,716){\special{em:lineto}}
\put(869,717){\special{em:lineto}}
\put(882,715){\special{em:lineto}}
\put(894,709){\special{em:lineto}}
\put(907,706){\special{em:lineto}}
\put(920,703){\special{em:lineto}}
\put(933,700){\special{em:lineto}}
\put(946,695){\special{em:lineto}}
\put(959,694){\special{em:lineto}}
\put(972,694){\special{em:lineto}}
\put(985,697){\special{em:lineto}}
\put(998,695){\special{em:lineto}}
\put(1011,696){\special{em:lineto}}
\put(1024,696){\special{em:lineto}}
\put(1037,697){\special{em:lineto}}
\put(1050,691){\special{em:lineto}}
\put(1063,688){\special{em:lineto}}
\put(1076,684){\special{em:lineto}}
\put(1089,682){\special{em:lineto}}
\put(1102,677){\special{em:lineto}}
\put(1115,674){\special{em:lineto}}
\put(1128,674){\special{em:lineto}}
\put(1141,676){\special{em:lineto}}
\put(1154,676){\special{em:lineto}}
\put(1167,675){\special{em:lineto}}
\put(1180,676){\special{em:lineto}}
\put(1193,677){\special{em:lineto}}
\put(1206,675){\special{em:lineto}}
\put(1219,669){\special{em:lineto}}
\put(1232,667){\special{em:lineto}}
\put(1245,664){\special{em:lineto}}
\put(1258,662){\special{em:lineto}}
\put(1271,656){\special{em:lineto}}
\put(1284,655){\special{em:lineto}}
\put(1297,655){\special{em:lineto}}
\put(1310,658){\special{em:lineto}}
\put(1323,656){\special{em:lineto}}
\put(1335,656){\special{em:lineto}}
\put(1348,657){\special{em:lineto}}
\put(1361,657){\special{em:lineto}}
\put(1374,654){\special{em:lineto}}
\put(1387,649){\special{em:lineto}}
\put(1400,646){\special{em:lineto}}
\put(1413,644){\special{em:lineto}}
\put(1426,641){\special{em:lineto}}
\sbox{\plotpoint}{\rule[-0.500pt]{1.000pt}{1.000pt}}%
\special{em:linewidth 1.0pt}%
\put(233,806){\usebox{\plotpoint}}
\put(233.00,806.00){\usebox{\plotpoint}}
\put(253.51,802.84){\usebox{\plotpoint}}
\multiput(259,802)(20.514,-3.156){0}{\usebox{\plotpoint}}
\put(274.03,799.69){\usebox{\plotpoint}}
\put(294.54,796.53){\usebox{\plotpoint}}
\multiput(298,796)(20.514,-3.156){0}{\usebox{\plotpoint}}
\put(315.06,793.38){\usebox{\plotpoint}}
\put(335.67,791.10){\usebox{\plotpoint}}
\multiput(337,791)(20.514,-3.156){0}{\usebox{\plotpoint}}
\put(356.20,788.05){\usebox{\plotpoint}}
\multiput(363,787)(20.514,-3.156){0}{\usebox{\plotpoint}}
\put(376.71,784.89){\usebox{\plotpoint}}
\put(397.23,781.73){\usebox{\plotpoint}}
\multiput(402,781)(20.514,-3.156){0}{\usebox{\plotpoint}}
\put(417.76,778.79){\usebox{\plotpoint}}
\put(438.37,776.40){\usebox{\plotpoint}}
\multiput(441,776)(20.473,-3.412){0}{\usebox{\plotpoint}}
\put(458.86,773.10){\usebox{\plotpoint}}
\multiput(466,772)(20.514,-3.156){0}{\usebox{\plotpoint}}
\put(479.37,769.94){\usebox{\plotpoint}}
\put(499.96,767.39){\usebox{\plotpoint}}
\multiput(505,767)(20.514,-3.156){0}{\usebox{\plotpoint}}
\put(520.51,764.61){\usebox{\plotpoint}}
\put(541.03,761.46){\usebox{\plotpoint}}
\multiput(544,761)(20.514,-3.156){0}{\usebox{\plotpoint}}
\put(561.58,758.65){\usebox{\plotpoint}}
\put(582.17,756.13){\usebox{\plotpoint}}
\multiput(583,756)(20.514,-3.156){0}{\usebox{\plotpoint}}
\put(602.68,752.97){\usebox{\plotpoint}}
\multiput(609,752)(20.694,-1.592){0}{\usebox{\plotpoint}}
\put(623.31,750.80){\usebox{\plotpoint}}
\put(643.83,747.64){\usebox{\plotpoint}}
\multiput(648,747)(20.514,-3.156){0}{\usebox{\plotpoint}}
\put(664.37,744.74){\usebox{\plotpoint}}
\put(684.97,742.31){\usebox{\plotpoint}}
\multiput(687,742)(20.514,-3.156){0}{\usebox{\plotpoint}}
\put(705.48,739.16){\usebox{\plotpoint}}
\multiput(713,738)(20.694,-1.592){0}{\usebox{\plotpoint}}
\put(726.11,736.98){\usebox{\plotpoint}}
\put(746.62,733.83){\usebox{\plotpoint}}
\multiput(752,733)(20.694,-1.592){0}{\usebox{\plotpoint}}
\put(767.25,731.65){\usebox{\plotpoint}}
\put(787.76,728.50){\usebox{\plotpoint}}
\multiput(791,728)(20.694,-1.592){0}{\usebox{\plotpoint}}
\put(808.39,726.32){\usebox{\plotpoint}}
\put(828.91,723.17){\usebox{\plotpoint}}
\multiput(830,723)(20.694,-1.592){0}{\usebox{\plotpoint}}
\put(849.53,720.99){\usebox{\plotpoint}}
\multiput(856,720)(20.514,-3.156){0}{\usebox{\plotpoint}}
\put(870.05,717.84){\usebox{\plotpoint}}
\put(890.63,715.28){\usebox{\plotpoint}}
\multiput(894,715)(20.514,-3.156){0}{\usebox{\plotpoint}}
\put(911.21,712.68){\usebox{\plotpoint}}
\put(931.80,710.18){\usebox{\plotpoint}}
\multiput(933,710)(20.514,-3.156){0}{\usebox{\plotpoint}}
\put(952.37,707.51){\usebox{\plotpoint}}
\multiput(959,707)(20.514,-3.156){0}{\usebox{\plotpoint}}
\put(972.94,704.85){\usebox{\plotpoint}}
\put(993.53,702.34){\usebox{\plotpoint}}
\multiput(998,702)(20.514,-3.156){0}{\usebox{\plotpoint}}
\put(1014.11,699.76){\usebox{\plotpoint}}
\put(1034.71,697.35){\usebox{\plotpoint}}
\multiput(1037,697)(20.514,-3.156){0}{\usebox{\plotpoint}}
\put(1055.27,694.59){\usebox{\plotpoint}}
\put(1075.85,692.02){\usebox{\plotpoint}}
\multiput(1076,692)(20.514,-3.156){0}{\usebox{\plotpoint}}
\put(1096.43,689.43){\usebox{\plotpoint}}
\multiput(1102,689)(20.514,-3.156){0}{\usebox{\plotpoint}}
\put(1117.01,686.85){\usebox{\plotpoint}}
\put(1137.62,684.52){\usebox{\plotpoint}}
\multiput(1141,684)(20.694,-1.592){0}{\usebox{\plotpoint}}
\put(1158.25,682.35){\usebox{\plotpoint}}
\put(1178.76,679.19){\usebox{\plotpoint}}
\multiput(1180,679)(20.694,-1.592){0}{\usebox{\plotpoint}}
\put(1199.39,677.02){\usebox{\plotpoint}}
\multiput(1206,676)(20.694,-1.592){0}{\usebox{\plotpoint}}
\put(1220.02,674.84){\usebox{\plotpoint}}
\put(1240.61,672.34){\usebox{\plotpoint}}
\multiput(1245,672)(20.514,-3.156){0}{\usebox{\plotpoint}}
\put(1261.19,669.75){\usebox{\plotpoint}}
\put(1281.79,667.34){\usebox{\plotpoint}}
\multiput(1284,667)(20.694,-1.592){0}{\usebox{\plotpoint}}
\put(1302.41,665.17){\usebox{\plotpoint}}
\multiput(1310,664)(20.694,-1.592){0}{\usebox{\plotpoint}}
\put(1323.04,662.99){\usebox{\plotpoint}}
\put(1343.61,660.34){\usebox{\plotpoint}}
\multiput(1348,660)(20.514,-3.156){0}{\usebox{\plotpoint}}
\put(1364.19,657.75){\usebox{\plotpoint}}
\put(1384.79,655.34){\usebox{\plotpoint}}
\multiput(1387,655)(20.694,-1.592){0}{\usebox{\plotpoint}}
\put(1405.41,653.17){\usebox{\plotpoint}}
\multiput(1413,652)(20.694,-1.592){0}{\usebox{\plotpoint}}
\put(1426,651){\usebox{\plotpoint}}
\sbox{\plotpoint}{\rule[-0.400pt]{0.800pt}{0.800pt}}%
\special{em:linewidth 0.8pt}%
\put(233,803){\special{em:moveto}}
\put(246,796){\special{em:lineto}}
\put(259,791){\special{em:lineto}}
\put(272,790){\special{em:lineto}}
\put(285,782){\special{em:lineto}}
\put(298,775){\special{em:lineto}}
\put(311,769){\special{em:lineto}}
\put(324,765){\special{em:lineto}}
\put(337,755){\special{em:lineto}}
\put(350,741){\special{em:lineto}}
\put(363,728){\special{em:lineto}}
\put(376,718){\special{em:lineto}}
\put(389,707){\special{em:lineto}}
\put(402,694){\special{em:lineto}}
\put(415,688){\special{em:lineto}}
\put(428,689){\special{em:lineto}}
\put(441,697){\special{em:lineto}}
\put(453,704){\special{em:lineto}}
\put(466,716){\special{em:lineto}}
\put(479,722){\special{em:lineto}}
\put(492,724){\special{em:lineto}}
\put(505,712){\special{em:lineto}}
\put(518,694){\special{em:lineto}}
\put(531,666){\special{em:lineto}}
\put(544,633){\special{em:lineto}}
\put(557,591){\special{em:lineto}}
\put(570,550){\special{em:lineto}}
\put(583,517){\special{em:lineto}}
\put(596,496){\special{em:lineto}}
\put(609,488){\special{em:lineto}}
\put(622,496){\special{em:lineto}}
\put(635,519){\special{em:lineto}}
\put(648,553){\special{em:lineto}}
\put(661,585){\special{em:lineto}}
\put(674,606){\special{em:lineto}}
\put(687,617){\special{em:lineto}}
\put(700,611){\special{em:lineto}}
\put(713,590){\special{em:lineto}}
\put(726,550){\special{em:lineto}}
\put(739,504){\special{em:lineto}}
\put(752,454){\special{em:lineto}}
\put(765,406){\special{em:lineto}}
\put(778,363){\special{em:lineto}}
\put(791,332){\special{em:lineto}}
\put(804,321){\special{em:lineto}}
\put(817,329){\special{em:lineto}}
\put(830,350){\special{em:lineto}}
\put(843,381){\special{em:lineto}}
\put(856,414){\special{em:lineto}}
\put(869,445){\special{em:lineto}}
\put(882,464){\special{em:lineto}}
\put(894,467){\special{em:lineto}}
\put(907,459){\special{em:lineto}}
\put(920,437){\special{em:lineto}}
\put(933,408){\special{em:lineto}}
\put(946,369){\special{em:lineto}}
\put(959,330){\special{em:lineto}}
\put(972,294){\special{em:lineto}}
\put(985,264){\special{em:lineto}}
\put(998,242){\special{em:lineto}}
\put(1011,231){\special{em:lineto}}
\put(1024,233){\special{em:lineto}}
\put(1037,246){\special{em:lineto}}
\put(1050,264){\special{em:lineto}}
\put(1063,286){\special{em:lineto}}
\put(1076,309){\special{em:lineto}}
\put(1089,329){\special{em:lineto}}
\put(1102,344){\special{em:lineto}}
\put(1115,350){\special{em:lineto}}
\put(1128,351){\special{em:lineto}}
\put(1141,343){\special{em:lineto}}
\put(1154,330){\special{em:lineto}}
\put(1167,310){\special{em:lineto}}
\put(1180,286){\special{em:lineto}}
\put(1193,262){\special{em:lineto}}
\put(1206,241){\special{em:lineto}}
\put(1219,221){\special{em:lineto}}
\put(1232,209){\special{em:lineto}}
\put(1245,206){\special{em:lineto}}
\put(1258,215){\special{em:lineto}}
\put(1271,231){\special{em:lineto}}
\put(1284,254){\special{em:lineto}}
\put(1297,282){\special{em:lineto}}
\put(1310,312){\special{em:lineto}}
\put(1323,340){\special{em:lineto}}
\put(1335,361){\special{em:lineto}}
\put(1348,376){\special{em:lineto}}
\put(1361,381){\special{em:lineto}}
\put(1374,376){\special{em:lineto}}
\put(1387,359){\special{em:lineto}}
\put(1400,335){\special{em:lineto}}
\put(1413,306){\special{em:lineto}}
\put(1426,280){\special{em:lineto}}
\sbox{\plotpoint}{\rule[-0.500pt]{1.000pt}{1.000pt}}%
\special{em:linewidth 1.0pt}%
\put(233,806){\usebox{\plotpoint}}
\put(233.00,806.00){\usebox{\plotpoint}}
\put(253.51,802.84){\usebox{\plotpoint}}
\multiput(259,802)(20.514,-3.156){0}{\usebox{\plotpoint}}
\put(274.03,799.69){\usebox{\plotpoint}}
\put(294.54,796.53){\usebox{\plotpoint}}
\multiput(298,796)(20.514,-3.156){0}{\usebox{\plotpoint}}
\put(315.06,793.38){\usebox{\plotpoint}}
\put(335.57,790.22){\usebox{\plotpoint}}
\multiput(337,790)(20.514,-3.156){0}{\usebox{\plotpoint}}
\put(356.08,787.06){\usebox{\plotpoint}}
\multiput(363,786)(20.694,-1.592){0}{\usebox{\plotpoint}}
\put(376.71,784.89){\usebox{\plotpoint}}
\put(397.23,781.73){\usebox{\plotpoint}}
\multiput(402,781)(20.514,-3.156){0}{\usebox{\plotpoint}}
\put(417.74,778.58){\usebox{\plotpoint}}
\put(438.25,775.42){\usebox{\plotpoint}}
\multiput(441,775)(20.473,-3.412){0}{\usebox{\plotpoint}}
\put(458.80,772.55){\usebox{\plotpoint}}
\multiput(466,772)(20.514,-3.156){0}{\usebox{\plotpoint}}
\put(479.37,769.94){\usebox{\plotpoint}}
\put(499.89,766.79){\usebox{\plotpoint}}
\multiput(505,766)(20.514,-3.156){0}{\usebox{\plotpoint}}
\put(520.40,763.63){\usebox{\plotpoint}}
\put(540.77,759.74){\usebox{\plotpoint}}
\multiput(544,759)(20.694,-1.592){0}{\usebox{\plotpoint}}
\put(561.36,757.33){\usebox{\plotpoint}}
\put(581.87,754.17){\usebox{\plotpoint}}
\multiput(583,754)(20.514,-3.156){0}{\usebox{\plotpoint}}
\put(602.38,751.02){\usebox{\plotpoint}}
\multiput(609,750)(20.514,-3.156){0}{\usebox{\plotpoint}}
\put(622.90,747.86){\usebox{\plotpoint}}
\put(643.49,745.35){\usebox{\plotpoint}}
\multiput(648,745)(20.224,-4.667){0}{\usebox{\plotpoint}}
\put(663.85,741.56){\usebox{\plotpoint}}
\put(684.37,738.41){\usebox{\plotpoint}}
\multiput(687,738)(20.514,-3.156){0}{\usebox{\plotpoint}}
\put(704.88,735.25){\usebox{\plotpoint}}
\put(725.40,732.09){\usebox{\plotpoint}}
\multiput(726,732)(20.514,-3.156){0}{\usebox{\plotpoint}}
\put(745.91,728.94){\usebox{\plotpoint}}
\multiput(752,728)(20.514,-3.156){0}{\usebox{\plotpoint}}
\put(766.42,725.78){\usebox{\plotpoint}}
\put(786.94,722.62){\usebox{\plotpoint}}
\multiput(791,722)(20.514,-3.156){0}{\usebox{\plotpoint}}
\put(807.45,719.47){\usebox{\plotpoint}}
\put(827.97,716.31){\usebox{\plotpoint}}
\multiput(830,716)(20.514,-3.156){0}{\usebox{\plotpoint}}
\put(848.48,713.16){\usebox{\plotpoint}}
\put(868.99,710.00){\usebox{\plotpoint}}
\multiput(869,710)(20.514,-3.156){0}{\usebox{\plotpoint}}
\put(889.49,706.75){\usebox{\plotpoint}}
\multiput(894,706)(20.514,-3.156){0}{\usebox{\plotpoint}}
\put(910.00,703.54){\usebox{\plotpoint}}
\put(930.51,700.38){\usebox{\plotpoint}}
\multiput(933,700)(20.224,-4.667){0}{\usebox{\plotpoint}}
\put(950.84,696.26){\usebox{\plotpoint}}
\put(971.35,693.10){\usebox{\plotpoint}}
\multiput(972,693)(20.694,-1.592){0}{\usebox{\plotpoint}}
\put(991.98,690.93){\usebox{\plotpoint}}
\multiput(998,690)(20.514,-3.156){0}{\usebox{\plotpoint}}
\put(1012.50,687.77){\usebox{\plotpoint}}
\put(1033.01,684.61){\usebox{\plotpoint}}
\multiput(1037,684)(20.514,-3.156){0}{\usebox{\plotpoint}}
\put(1053.52,681.46){\usebox{\plotpoint}}
\put(1074.04,678.30){\usebox{\plotpoint}}
\multiput(1076,678)(20.514,-3.156){0}{\usebox{\plotpoint}}
\put(1094.55,675.15){\usebox{\plotpoint}}
\multiput(1102,674)(20.514,-3.156){0}{\usebox{\plotpoint}}
\put(1115.07,671.99){\usebox{\plotpoint}}
\put(1135.58,668.83){\usebox{\plotpoint}}
\multiput(1141,668)(20.514,-3.156){0}{\usebox{\plotpoint}}
\put(1156.10,665.68){\usebox{\plotpoint}}
\put(1176.61,662.52){\usebox{\plotpoint}}
\multiput(1180,662)(20.514,-3.156){0}{\usebox{\plotpoint}}
\put(1197.12,659.37){\usebox{\plotpoint}}
\put(1217.64,656.21){\usebox{\plotpoint}}
\multiput(1219,656)(20.514,-3.156){0}{\usebox{\plotpoint}}
\put(1238.15,653.05){\usebox{\plotpoint}}
\multiput(1245,652)(20.514,-3.156){0}{\usebox{\plotpoint}}
\put(1258.67,649.95){\usebox{\plotpoint}}
\put(1279.29,647.72){\usebox{\plotpoint}}
\multiput(1284,647)(20.514,-3.156){0}{\usebox{\plotpoint}}
\put(1299.81,644.57){\usebox{\plotpoint}}
\put(1320.32,641.41){\usebox{\plotpoint}}
\multiput(1323,641)(20.473,-3.412){0}{\usebox{\plotpoint}}
\put(1340.81,638.11){\usebox{\plotpoint}}
\multiput(1348,637)(20.514,-3.156){0}{\usebox{\plotpoint}}
\put(1361.33,634.95){\usebox{\plotpoint}}
\put(1381.84,631.79){\usebox{\plotpoint}}
\multiput(1387,631)(20.694,-1.592){0}{\usebox{\plotpoint}}
\put(1402.47,629.62){\usebox{\plotpoint}}
\put(1422.98,626.46){\usebox{\plotpoint}}
\end{picture}

\vskip 15pt

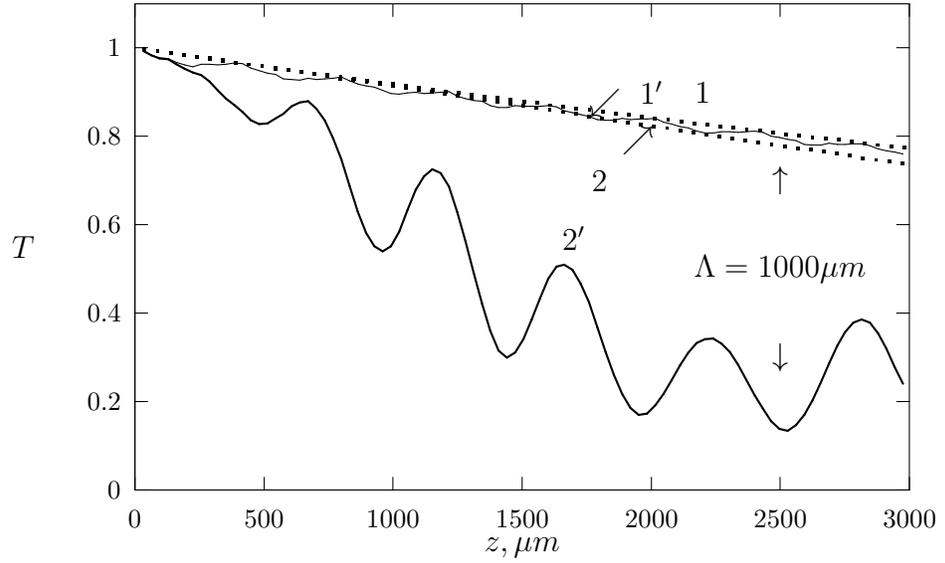
\centerline{\bf Fig.~5}
\noindent {\small 
The evolution of total intensity (curves $1$ and $2$, point) and basic mode "0"
(curves $1'$ and $2'$) of X-ray beam with energy
$E=17keV$ in the channel Cr/C/Cr, $d=1620 \AA$ 
for direct channel ($1$ and $1'$) and under deformations with the period $\Lambda=1000\mu m$ 
($2$ and $2'$).
$\sigma = 0 \AA$, $L=3mm$,  $a= 120 \AA$.
The initial beam corresponds to the basic waveguide mode.
}
\end{minipage}
\end{center}
\vskip 20pt


\begin{thebibliography}{99}
%\begin{references}

\bibitem{Riekel2000} {\sc C.~Riekel},
Report Progress Phys., {\bf 63} (2000), 232.

\bibitem{Jark1999p9} {\sc W.~Jark, S.~Di Fonzo, G.~Soullie, 
A.~Cedola,
S.~Lagomarsino}, J. Alloys and Compounds,
{\bf 286} (1999), 9-13.

\bibitem{Kukhlevsky1999pure} {\sc S.V.~Kukhlevsky, G.~Lubkovics,
K.~Negrea, L.~Kozma},
Pure Appl. Opt., {\bf 6} (1999), 97.
 
\bibitem{Kukhlevsky2000} {\sc S.V.~Kukhlevsky, F.~Flora, 
A.~Marinai, G.~Nyitray,
Zs.~Kozma, A.~Ritucci, L.~Palladino, A.~Reale, G.~Tomassetti},
X-ray spectrometry, {\bf 29} (2000), 0000.

\bibitem{Kantsyrev1995} {\sc V.L.~Kantsyrev, R.~Bruch, 
M.~Bailey, A.~Shlaptseva}, Applied Phys. Lett.,
{\bf 66}, n.26 (1995), 3567.

\bibitem{Fanchenko1999} {\sc S.S.~Fanchenko, A.A.~Nefedov}, 
phys. stat. solidi (b),
{\bf 212/1} (1999), R3.

\bibitem{Vinogradov1985} {\sc A.V.~Vinogradov, N.N.~Zorev, 
I.V.~Kozhevnikov, I.G.~Yakushkin},
Sov. Phys. JETP, {\bf 62} (1985), 1225.

%\bibitem{28} {\sc A.V.~Andreev}, Uspehi Fiz. Nauk, {\bf 145},
%113 (1985).


\bibitem{BobrovaJEPT1999} {\sc T.A.~Bobrova, L.I.~Ognev}, 
JEPT Letters, {\bf 69} (1999), 734.


\bibitem{BobrovaPhysStatSol1997} {\sc T.A.~Bobrova, L.I.~Ognev},
phys. stat. sol. (b), {\bf 203/2} (1997), R11.


\bibitem{Tsuji1995} {\sc K.~Tsuji, T.~Yamada, H.~Hirokava}, 
J. Applied Phys., {\bf 78} (1995), 969.

\bibitem{Henke1993} {\sc B.L.~Henke,~E.M.~Gullikson, 
J.C.~Davis}, Atomic Data and Nuclear Data 
\newline
Tables, {\bf 54}, no. 2 (1993), 181-342.
(Excess 
\newline 
http://www-cxro.lbl.gov/optical\_constants/).


\bibitem{HolyPhysStatSol1987} {\sc V.~Hol\'y, K.T.Gabrielyan}, 
phys. stat. sol. (b), {\bf 140} (1987), 39.

\bibitem{OgnevREffDS1993} {\sc L.I.~Ognev}, Radiation Effects  
and  Defects  in  Solids, {\bf 25} (1993), 81.

%\bibitem{18} {\sc L.I.~Ognev}, Nucl. Instrum. Meth. in Phys. 
%Res., {\bf B84}, 319 (1994).

\bibitem{BobrovaPrep1997} {\sc T.A.~Bobrova, L.I.~Ognev}, 
Preprint IAE-6051/11, Moscow, 1997 (in Russian; 
English translation
can be obtained from http://xxx.itep.ru/abs/physics/9807033).

\bibitem{OgnevTPL2000} {\sc L.I.~Ognev}, Technical Phys. Lett.,
{\bf 26} (2000), 67-69.

\bibitem{Jark_priv} {\sc W.~Jark}, private communication.

%\end{references}
\end{thebibliography}
\end{document}